\documentclass[aps,prb,twocolumn,showpacs,10pt,floatfix]{revtex4-2}

\usepackage{natbib}
\usepackage{amsfonts}
\usepackage{amsmath}
\usepackage{amssymb}
\usepackage{braket}
\usepackage{mathtools}
\usepackage{rotating}
\usepackage{color}
\usepackage{framed}
\usepackage{enumerate}
\usepackage{graphicx}
\usepackage{float}
\usepackage{hyperref}
\hypersetup{
    colorlinks=true,
    linkcolor=blue,    
    urlcolor=blue,
    citecolor=blue
    }
\usepackage{tikz}
\usepackage{circuitikz}
\usepackage{tabularx}
\usepackage{multirow}
\usepackage{csquotes}

\definecolor{blue-plot}{rgb}{0, 0.4, 0.8}
\definecolor{orange-plot}{rgb}{1, 0.6, 0}

\definecolor{blue-1}{rgb}{0.776,0.832,0.872}
\definecolor{blue-2}{rgb}{0.746,0.802,0.842}
\definecolor{blue-3}{rgb}{0.806,0.862,0.902}
\definecolor{gray-1}{rgb}{0.88,0.892,0.896}
\definecolor{gray-2}{rgb}{0.85,0.862,0.866}
\definecolor{gray-3}{rgb}{0.91,0.922,0.926}

\newcommand{\vm}{\vec m}
\newcommand{\vE}{\vec E}

\newcommand{\vj}{\vec j}
\newcommand{\vi}{\vec i}

\newcommand{\ve}{\vec e}
\newcommand{\vmu}{\mbox{\boldmath $\mu$}}

\newcommand{\vjs}{{\vec j}_{\rm s}}

\newcommand{\vmus}{\vmu_{\rm s}}
\renewcommand{\vec}[1]{\mathbf{#1}}
\newcommand{\Dex}{D_{\rm ex}}

\begin{document}

\title{Resonant current-in-plane spin-torque diode effect in magnet--normal metal bilayers}

\author{Ulli Gems}
\author{Oliver Franke}
\author{Piet W. Brouwer}

\affiliation{Dahlem Center for Complex Quantum Systems and Physics Department, Freie Universit\"at Berlin, Arnimallee 14, 14195 Berlin, Germany}

\begin{abstract}
Via the spin-Hall effect and its inverse, in-plane charge currents in a normal metal--ferromagnet (N$|$F) bilayer can be used to excite and detect magnetization dynamics in F. Using a magneto-electric circuit approach, we here consider the current response to quadratic order in the applied electric field, which is resonantly enhanced for driving frequencies close to frequencies of coherent magnetization modes. Our theory can be applied to bilayers with a magnetic insulator or with a magnetic metal. It focuses on the contribution of coherent magnetization dynamics to spin currents collinear with the equilibrium magnetization direction, but also takes into account relaxation of spin accumulation via spin currents carried by incoherent magnons and conduction electrons in F.
\end{abstract}

\maketitle

\newpage

{\em Introduction.---}
The spin-Hall magnetoresistance effect (SMR), a correction to the in-plane conductivity of a normal metal--ferromagnet (N$|$F) bilayer that depends on the magnetization direction of F, is one of the central spintronic phenomena \cite{Weiler2012-gh,Huang2012-fu,Nakayama2013-gf,Hahn2013-rw,Vlietstra2013-uv,Althammer2013-zm,Chen2013-gf,Chen2016-pc}.
Via the spin-Hall effect \cite{Dyakonov1971-pp,Hirsch1999-gc}, the in-plane applied electric field drives a spin current through the N$|$F interface, which, via the inverse spin-Hall effect, causes a correction to the in-plane charge current in N. 
Since the spin current from N to F depends on the direction of the magnetization in F, the correction to the charge current also depends on it. 
The origin for the dependence on the (equilibrium) magnetization direction $\vm^{\rm eq}$ is that spin currents polarized collinear with and perpendicular to $\vm^{\rm eq}$ have fundamentally different mechanisms and, hence, different magnitude:
Conduction electrons in a metallic ferromagnet and incoherent, thermal magnons carry a ``longitudinal'' spin current \cite{Xiao2010-wi,Bender2015-tr,Zhang2019-zv}, polarized along the magnetization direction, whereas a ``transverse'' spin current perpendicular to $\vm^{\rm eq}$ comes from coherent magnetization dynamics, driven by the current-induced spin torque at the N$|$F interface \cite{Tserkovnyak2005-dt}. 

If the frequency $\omega$ of the applied electric field $\vE$ matches that of a magnetization mode in F, such a mode will be resonantly driven by the spin currents from the spin-Hall effect, an effect known as spin-torque ferromagnetic resonance \cite{Liu2011-sy,Kondou2012-kt,Chiba2014-mh,Schreier2015-pt,Sklenar2015-ab,He2016-ab}. To electrically detect the spin-torque ferromagnetic resonance, one makes use of the fact that there also is a longitudinal spin current associated with coherent magnetization precession, which is quadratic in the magnetization amplitude \cite{Tserkovnyak2002-ax} and, hence, gives a charge current response at frequencies $\Omega = 0$ and $\Omega = 2 \omega$ \cite{Ganguly2014-zw}.
The rectified response at $\Omega = 0$ is the in-plane counterpart of the current-perpendicular-to-plane spin-torque-diode effect in F$|$N$|$F trilayers \cite{Tulapurkar2005-ew,Sankey2006-cf}.

\begin{figure}
    \centering
    \includegraphics[width=1.0\linewidth]{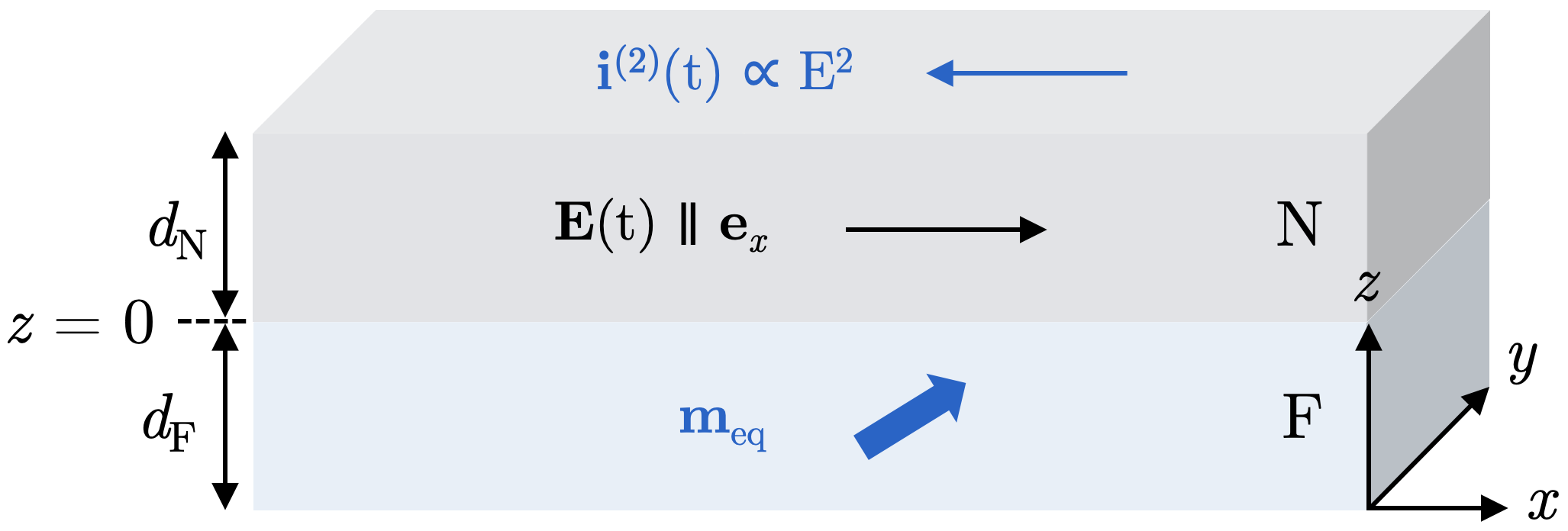}
    \caption{Geometry of the N$|$F bilayer, consisting of a normal metals N and a metallic or insulating ferromagnet F. An in-plane electric field $\vE(t) = E(e^{-i \omega t} + e^{i \omega t}) \ve_x$ in N drives a coherent magnetization mode in F. In this article, we calculate the quadratic-in-$E$ current response $\vi_{i}^{(2)}$ at frequencies $\Omega = 0$ and $\Omega = 2 \omega$ that arises from the modulation of the spin-Hall magnetoconductance by the precessing magnetization in F.}
\label{fig:geometry}
\end{figure}

A theory of the resonant in-plane spin-torque diode effect was developed by Chiba, Bauer, and Takahashi \cite{Chiba2014-mh} (see also Ref.\ \onlinecite{Chiba2015-jt}). These authors considered a bilayer consisting of a normal metal with spin-Hall effect (such as Pt) and a magnetic insulator (such as YIG), accounting for the spin currents through the N$|$F interface associated with the coherent magnetization dynamics in F \cite{Tserkovnyak2005-dt}.
In the present article, we present a calculation of the resonant in-plane spin-torque diode effect in which we also include longitudinal spin transport by conduction electrons and incoherent magnons \cite{Bender2015-tr,Zhang2019-zv,Reiss2022-tm}. Although longitudinal spin currents do not resonantly depend on the driving frequency $\omega$ \cite{Reiss2021-em}, the existence of these additional longitudinal spin transport channels across the N$|$F interface quantitatively affects the quadratic-in-$E$ charge current associated with the longitudinal spin current from coherent magnetization dynamics. The inclusion of longitudinal spin currents is especially relevant for bilayers involving a ferromagnetic metal (instead of the ferromagnetic insulator considered in Ref.\ \onlinecite{Chiba2014-mh}), because there the large longitudinal spin current carried by conduction electrons dominates the interfacial spin transport.

To adequately account for the different contributions to longitudinal and transverse spin transport, we make use of a magneto-electric circuit picture \cite{Roy2021-ab,Reiss2021-em}. We also consider the magnetization beyond the macrospin approximation (while still assuming in-plane translation invariance for the coherent magnetization dynamics), so that not only resonances at the uniform ferromagnetic resonance frequency, but also at higher magnon frequencies are included in our calculation.

The remainder of this article is organized as follows: After a brief introduction to the system and our notation, we first consider the linear-in-field spin-Hall magnetoresistance. Closely following Ref.\ \onlinecite{Reiss2021-em}, we formulate the relations between spin accumulation, magnetization amplitude, spin currents, and other relevant variables in terms of a magneto-electric circuit. We apply the circuit equations at the frequency $\omega$ of the driving field $\vE$ to find the linear-in-$E$ response, but we also need the circuit equations at the frequencies $\Omega = 0$ and $\Omega = 2 \omega$ to find the quadratic-in-$E$ response.
We numerically evaluate the diode effect for typical device parameters of a Pt$|$YIG bilayer and for an Au$|$Fe bilayer and close with a brief discussion.

{\em System and notation.---} We consider an F$|$N bilayer geometry, shown schematically in Fig.\ \ref{fig:geometry}. We choose coordinates such that the $z$ direction is perpendicular to the layers, the normal metal N of thickness $d_{\rm N}$ is located at $0 < z < d_{\rm N}$, and the magnet F at $-d_{\rm F} < z < 0$. A spatially uniform time-dependent electric field 
\begin{equation}
  \vE(t) = E (e^{-i \omega t} + e^{i \omega t}) \ve_x
\end{equation}
is applied in N. The linear-in-$E$ charge current response is of the form 
\begin{equation}
 \bar i^{x/y}(t) = \bar i^{x/y}_{\omega} e^{-i \omega t} + \bar i^{x/y}_{-\omega} e^{i \omega t}
  \label{eq:i1}
\end{equation}
and was considered in Refs.\ \onlinecite{Chen2013-gf,Chen2016-pc,Zhang2019-zv,Reiss2021-em}. In this article, we are interested in the quadratic-in-$E$ average charge current $\bar i^{x/y(2)}(t)$,
\begin{equation}
  \bar i^{x/y(2)}(t) = \bar i^{x/y(2)}_{2 \omega} e^{-2 i \omega t} + \bar i^{x/y(2)}_{{\rm c},0}
  + \bar i^{x/y(2)}_{-2\omega} e^{2 i \omega t}.
  \label{eq:i2}
\end{equation}
For the other relevant variables, such as the spin current density through the N$|$F interface $\vjs^z$ and the interfacial spin accumulation $\vmus$ in N \footnote{We define the spin current density as $j_{{\rm s}z}^z = (\hbar/2e)[j_{{\rm c}\uparrow}^z - j_{{\rm c}\downarrow}^z]$, where $j_{{\rm c}\uparrow,\downarrow}^z$ is the charge current density carried by electrons with spin $\uparrow, \downarrow$ projected onto the $z$ axis, respectively. Similarly, the spin accumulation $\mu_{{\rm s}z} = \mu_{\uparrow} - \mu_{\downarrow}$, where $\mu_{\uparrow,\downarrow}$ is the (electro-)chemical potential for electrons with spin $\uparrow$, $\downarrow$ projected onto the $z$ axis. We use analogous definitions for the components $j_{{\rm s}x}^{z}$, $j_{{\rm s}y}^z$, $\mu_{{\rm s}x}$, and $\mu_{{\rm s}y}$.}, we also write their time dependence in terms of Fourier components at frequencies $-\omega$ and $\omega$ for the linear response and at frequencies $-2 \omega$, $0$, and $2 \omega$ for the quadratic-in-$E$ response.

{\em Longitudinal and transverse spin transport.---} To describe spin transport across the N$|$F interface and in F, we choose a separate set of coordinate axes $\ve_1$, $\ve_2$, and $\ve_3$, such that $\ve_3 = \vm^{\rm eq}$ points along the equilibrium magnetization direction in F. We find it advantageous to combine $\ve_1$ and $\ve_2$ into the complex unit vector 
\begin{equation}
  \ve_{\perp} = \frac{1}{\sqrt{2}}(\ve_1 + i \ve_2).
\end{equation}
The magnetization direction at the F$|$N interface $\vm$ can then be written as
\begin{equation}
  \vm = m_{\perp} \ve_{\perp} + m_{\perp}^* \ve_{\perp}^* +
  \vm^{\rm eq} \sqrt{1 - 2 |m_{\perp}|^2}.
  \label{eq:m1}
\end{equation}
To account for the different response to spin excitations collinear with and perpendicular to the equilibrium magnetization direction $\vm^{\rm eq}$, we separate the spin accumulation $\vmus$ into ``longitudinal'' and ``transverse'' components with respect to $\vm^{\rm eq}$,
\begin{align}
  \vmus =&\, 
  \mu_{{\rm s}\perp} \ve_{\perp} + \mu^*_{{\rm s}\perp} \ve^*_{\perp}
  + \mu_{{\rm s}\parallel}\vm^{\rm eq} .
  \label{eq:decomposition_long_trans}
\end{align}
We use an analogous decomposition into longitudinal and transverse components for the spin current $\vjs$ at the N$|$F interface.

{\em Spin-Hall effect and inverse spin-Hall effect.---} An electric field in N causes a spin accumulation $\vmus$ and a spin current $\vjs^z$ at the N$|$F interface \cite{Dyakonov1971-pp,Dyakonov1971-yh,Hirsch1999-gc,Takahashi2006-tz}. 
To linear order in $E$, the resulting spin accumulation and spin current at the N$|$F interface satisfy the relation \cite{Chen2013-gf,Chen2016-pc,Chiba2014-mh,Reiss2021-em}
\begin{align}
  \label{eq:nmspinrelation}
  Z_{{\rm N}} \vj^z_{{\rm s},\omega} =
  \vmu_{{\rm s},\omega} -  2 e
  \lambda_{{\rm N}} \theta_{{\rm SH}} E \ve_y,
\end{align}
where $\theta_{\rm SH}$ is the phenomenological spin-Hall angle, $\lambda_{\rm N}$ the spin diffusion length in N, and $Z_{\rm N}$ is the ``spin impedance'' of N \cite{Reiss2021-em},
\begin{equation}
  Z_{{\rm N}} = \frac{2 e^2}{\hbar}
  \frac{2 \lambda_{{\rm N}}}{\sigma_{{\rm N}}}.
\label{eq:zn}
\end{equation}
In Eq.~(\ref{eq:zn}), $\sigma_{\rm N}$ is the conductivity of N and it was assumed that $\lambda_{\rm N} \ll d_{\rm N}$. Conversely, a spin accumulation $\vmus$ at the N$|$F interface gives rise to a charge current $\delta \bar i^{x/y}$ (averaged over the width $d_{\rm N}$ of N) in the normal metal N \cite{Chen2013-gf,Chen2016-pc,Reiss2021-em},
\begin{align}
  \label{eq:deltaix}
  \delta \bar i^{x}(\omega) =&\ \theta_{{\rm SH}}
     \frac{  \sigma_{{\rm N}}}{2 d_{{\rm N}} e} 
     \mu_{{\rm s}y}(\omega), \\
  \label{eq:deltaiy}
  \delta \bar i^{y}(\omega) =&\, - \theta_{{\rm SH}}
     \frac{\sigma_{{\rm N}}}{2 d_{{\rm N}} e} 
    \mu_{{\rm s}x}(\omega).
\end{align}

\begin{figure}
    \centering
    \includegraphics[width=1.0\linewidth]{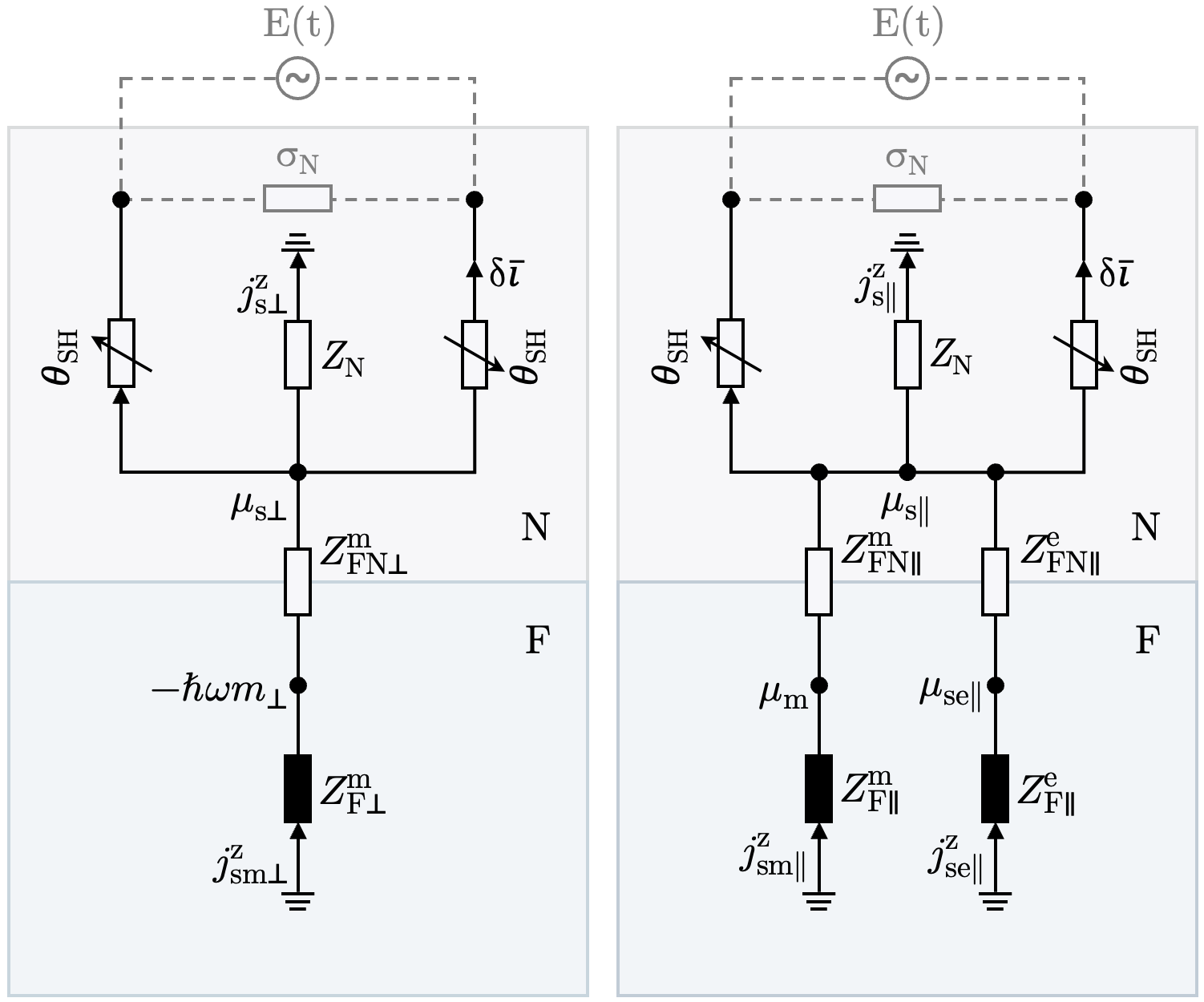}
    \caption{Equivalent magneto-electronic circuit diagrams for the linear charge and spin current response of an N$|$F bilayer to an applied electric field in N. The left and right circuits show the corrections arising from spin currents and spin accumulations perpendicular to and collinear with the magnetization direction in F, respectively. Dashed and solid lines indicate transport of charge and spin, respectively. The relations between the spin and charge currents and the generalized potentials $\mu_{{\rm se}\parallel,\omega}$, $\mu_{{\rm m},\omega}$, and $-\hbar \omega m_{\perp,\omega}$ are given in Eqs.~(\ref{eq:nmspinrelation})--(\ref{eq:fspinperp}). The ``grounds'' in the diagram represent the fact that spin is not a conserved quantity in N and in F.}
\label{fig:circuits}
\end{figure}

{\em Linear response and magneto-electric circuit theory.---} To complete the calculation of the charge response, it remains to find a relation between the interfacial spin accumulation in N $\vmus$ and the spin current $\vjs$ at the N$|$F interface. 
The linear-in-$E$ spin current $\vj_{{\rm s},\omega}^z$ at the F$|$N interface is the sum of a longitudinal contribution from conduction electrons and incoherent magnons in F and a transverse contribution from the coherent magnetization dynamics,
\begin{equation}
  \label{eq:jconservation}
  \vj_{{\rm s},\omega}^z =
  \left( 
  j_{{\rm se}\parallel,\omega}^z 
  + j_{{\rm sm}\parallel,\omega}^z 
  \right)
  \vm^{\rm eq} +
  j_{{\rm sm}\perp,\omega}^z \ve_{\perp} +
  j_{{\rm sm}\perp,\omega}^{z*} \ve_{\perp}^*.
\end{equation}
The three individual contributions to the spin current can be expressed in terms of the difference of the appropriate longitudinal or transverse component of the spin accumulation $\vmu_{{\rm s},\omega}$ in N and the spin accumulation $\mu_{{\rm se}\parallel,\omega}$ in F, the magnon chemical potential $\mu_{{\rm m},\omega}$ in F, and the complex magnetization amplitude $m_{\perp}$, and the corresponding interface spin impedances $Z_{{\rm FN}\parallel}^{\rm e}$, $Z_{{\rm FN}\parallel}^{\rm m}$, and  $Z_{{\rm FN}\perp}^{\rm m}$, respectively \cite{Reiss2021-em},
\begin{align}
  \label{eq:fnspinrelationel}
  Z_{{\rm FN}\parallel}^{\rm e} j_{{\rm se}\parallel,\omega} =&\,
  - [\mu_{{\rm s}\parallel,\omega} -\mu_{{\rm se}\parallel,\omega}], \\
  \label{eq:fnspinrelationth}
  Z_{{\rm FN}\parallel}^{\rm m} j_{{\rm sm}\parallel,\omega} =&\,
  - [\mu_{{\rm s}\parallel,\omega} - \mu_{{\rm sm}\parallel,\omega}], \\
  Z_{{\rm FN}\perp}^{\rm m} j_{{\rm sm}\perp,\omega} =&\,
  - [\mu_{{\rm s}\perp,\omega} + \hbar \omega m_{\perp,\omega}].
\label{eq:fnspinrelationperp}
\end{align}
Solving the equations of motion for the conduction electrons in F, incoherent magnons in F, and the coherent magnetization dynamics yields additional equations relating spin currents and $\mu_{{\rm se}\parallel,\omega}$, $\mu_{{\rm m},\omega}$, and $m_{\perp,\omega}$,
\begin{align}
  Z_{{\rm F}\parallel}^{\rm e}
  j_{{\rm se}\parallel,\omega} =&\, - \mu_{\rm se,\omega} \\
  Z_{{\rm F}\parallel}^{\rm m}(\omega)
  j_{{\rm sm}\parallel,\omega} =&\, - \mu_{{\rm m},\omega}, \\
  Z_{{\rm F}\perp}^{\rm m}(\omega)
  j_{{\rm sm}\perp,\omega} =&\ \hbar \omega m_{\perp}(\omega),
  \label{eq:fspinperp}
\end{align}
where $Z_{{\rm F}\parallel}^{\rm e}$, $Z_{{\rm F}\parallel}^{\rm m}(\omega)$, and $Z_{{\rm F}\perp}^{\rm m}(\omega)$ are spin impedances for the conduction electrons, thermal magnons, and coherent magnetization dynamics in F, respectively. Equations (\ref{eq:nmspinrelation})--(\ref{eq:fspinperp}) form a closed set, which is represented schematically in the circuit diagrams of Fig.\ \ref{fig:circuits}. Here, we only need the solution of these linear-response equations for the transverse spin current, which reads
\begin{align}
  \label{eq:jsolution}
  j^z_{{\rm sm}\perp,\omega} =&\, 
  - \frac{\hbar}{2 e} \theta_{\rm SH} \sigma_{\rm N} \frac{Z_{\rm N}}{Z_{\perp}(\omega)} E \ve_{\perp}^* \cdot \ve_y,
\end{align}
where $Z_{\perp}(\omega)$ is the transverse series impedance,
\begin{equation}
  Z_{\perp}(\omega) = Z_{\rm N} + Z_{{\rm FN}\perp}^{\rm m} + Z_{{\rm F}\perp}^{\rm m}(\omega).
\end{equation}

Explicit expressions for the four longitudinal impedances $Z_{{\rm FN}\parallel}^{\rm e}$, $Z_{{\rm FN}\parallel}^{\rm m}$, $Z_{{\rm F}\parallel}^{\rm e}$, and $Z_{{\rm F}\parallel}^{\rm m}(\omega)$ in terms of the properties of the F$|$N interface and F are given in Ref.\ \onlinecite{Reiss2021-em} and in the appendix. Of these, only $Z_{{\rm F}\parallel}^{\rm m}(\omega)$ has an appreciable (but non-resonant) frequency dependence for frequencies up to the THz range. Below, we discuss the two impedances $Z_{{\rm FN}\perp}^{\rm m}$ and $Z_{{\rm F}\perp}^{\rm m}(\omega)$ for the transverse response in more detail, since they are key to the resonant nonlinear response. 

{\em Transverse spin transport in F.---} Spin currents with polarization transverse to the magnetization direction arise from the coherent magnetization dynamics in F. We describe the magnetization dynamics using the Landau-Lifshitz-Gilbert equation \cite{Gilbert2004-gi,Landau1980-dj}
\begin{align}
  \label{eq:llg}
  \dot \vm(z) =&\ \omega_0 \vm^{\rm eq} \times \vm(z) + \alpha \vm(z) \times \dot \vm(z)
 \nonumber \\ &\, \mbox{}  + \frac{1}{\hbar s} \frac{\partial}{\partial z} \vj_{{\rm ms}}^{z}(z),
\end{align}
where $\omega_0$ is the ferromagnetic-resonance frequency, which includes effects of static external magnetic fields, demagnetization field, and anisotropies, $\alpha$ is the bulk Gilbert damping coefficient, and $s = M_{\rm s}/\gamma$ the spin per unit volume (with $M_{\rm s}$ the magnetic moment per unit volume and $\gamma$ the gyromagnetic ratio). The spin current density from the magnetization dynamics is \cite{Barnes2006-wr}
\begin{equation}
  \vj_{{\rm sm}}^z(z) = - \hbar \Dex s \vm(z) \times \frac{\partial \vm(z)}{\partial z},
  \label{eq:jsllg}
\end{equation}
where $\Dex$ is the spin stiffness of F. 

Inserting the parameterization (\ref{eq:m1}) and keeping terms to linear order in $m_{\perp}(z)$ only, one gets the linearized equations
\begin{align}
  j_{{\rm sm}\perp,\omega}^z(z) =&\
  i \hbar s \Dex \frac{\partial m_{\perp,\omega}(z)}{\partial z},
  \label{eq:jlin} \\
  \label{eq:mlin}
  \frac{\partial}{\partial z}
  j_{{\rm sm}\perp,\omega}^z(z) =&\,
  -i \hbar s (\omega + i \alpha \omega - \omega_0) m_{\perp,\omega}(z).
\end{align}
Solving Eqs.~(\ref{eq:jlin}) and (\ref{eq:mlin}) with the boundary condition that the spin current must vanish at the boundary of the F layer at $z=-d_{\rm F}$, we find that the magnetization amplitude $m_{\perp} \equiv m_{\perp}(0)$ at the F$|$N interface and the interfacial spin current $j_{{\rm sm}\perp,\omega}^z \equiv j_{{\rm sm}\perp,\omega}^z(0)$ satisfy a relation of the form (\ref{eq:fspinperp}), with
\begin{equation}
  \label{eq:ZF2}
  Z_{{\rm F}\perp}^{\rm m}(\omega) =  
  \frac{i \hbar \omega}{\Dex s k(\omega)} 
  \cot[k(\omega) d_{\rm F}].
\end{equation}
Here $k(\omega)$ is the (complex) wavenumber of the coherent magnetization mode at frequency $\omega$, which is the solution of
\begin{equation}
  \omega_0 + \Dex k^2 = \omega(1 + i \alpha).
  \label{eq:k}
\end{equation}
The spin impedance $Z_{{\rm F}\perp}^{\rm m}(\omega)$ has sharp peaks near the resonance frequencies $\omega = \omega_n$ with
\begin{equation}
  \omega_n = \Dex \left( \frac{n \pi}{d_{\rm F}} \right)^2 + \omega_0.
  \label{eq:omegan}
\end{equation}

{\em Transverse spin transport through the F$|$N interface.---} The spin current through the interface that is associated with the coherent magnetization dynamics is a sum of spin-torque and spin-pumping contributions \cite{Tserkovnyak2005-dt}
\begin{equation}
  \vj^z_{{\rm sm}} =
  \frac{1}{4 \pi} \left( \mbox{Re}\, g_{\uparrow\downarrow}
    \vm \times + \mbox{Im}\, g_{\uparrow\downarrow} \right)
    \left(\vm \times \vmu_{\rm s} + \hbar \dot \vm \right),
  \label{eq:jsinterface}
\end{equation}
where $g_{\uparrow\downarrow}$ is the spin-mixing conductance of the interface  \cite{Brataas2000-ar,Xia2002-xc}.
To linear order in $\vmus$ and $\dot \vm$, the spin current associated with the coherent magnetization dynamics is purely transverse, $\vj^z_{{\rm sm}} = j^z_{{\rm sm}\perp} \ve_{\perp} + j^{z*}_{{\rm sm}\perp} \ve_{\perp}^*$, and one finds that $j^z_{{\rm sm}\perp,\omega}$ and $\mu_{{\rm s}\perp,\omega} + \hbar \omega m_{\perp,\omega}$ are related by an equation of the form (\ref{eq:fnspinrelationperp}) with $Z_{{\rm FN}\perp}^{\rm m}$ given by \cite{Reiss2021-em}
\begin{equation}
  Z_{{\rm FN}\perp}^{\rm m} = \frac{4 \pi}{g_{\uparrow\downarrow}}.
\end{equation}

{\em Quadratic response from longitudinal spin current.---}
To find the leading contribution to the longitudinal spin current associated with the coherent magnetization dynamics, we must expand Eq.~(\ref{eq:jsinterface}) to second order in $m_{\perp}$,
\begin{equation}
  \label{eq:j2long}
  j^{z(2)}_{{\rm sm}\parallel}(t) = - [m_{\perp}^*(t) j^z_{{\rm sm}\perp}(t)
  + m_{\perp}(t) j^z_{{\rm sm}\perp}(t)^*].
\end{equation}
Fourier transforming to time and eliminating $m_{\perp,\omega}^*$ using Eq.~(\ref{eq:fspinperp}), we find the dc contributions to the second-order-in-$E$ longitudinal spin current,
\begin{align}
  \label{eq:j20}
\begin{split}
  j^{z(2)}_{{\rm sm}\parallel,0} =
  - \frac{2}{\hbar \omega} 
  \big[ &|j_{{\rm sm}\perp,\omega}^z|^2
  \mbox{Re}\, Z_{{\rm F}\perp}^{\rm m}(\omega)  \\
  &+ |j_{{\rm sm}\perp,-\omega}^z|^2
  \mbox{Re}\, Z_{{\rm F}\perp}^{\rm m}(-\omega) \big],
\end{split}
\end{align}
and the contribution at frequency $\Omega = \pm 2 \omega$,
\begin{align}
  \label{eq:j22}
\begin{split}
  j^{z(2)}_{{\rm sm}\parallel,\pm 2 \omega} =&\,
  - \frac{1}{\hbar \omega}
  j_{{\rm sm}\perp,\pm \omega}^z j_{{\rm sm}\perp,\mp \omega}^{z*} \\ &\, \mbox{} \times
  [Z_{{\rm F}\perp}^{\rm m}(\mp \omega)^* + Z_{{\rm F}\perp}^{\rm m}(\pm \omega)]. 
\end{split}
\end{align}
The linear-in-$E$ transverse spin currents $j_{{\rm sm}\perp,\omega}^z$ and $j_{{\rm sm}\perp,- \omega}$ that appear on the right-hand-side of Eqs.~(\ref{eq:j20}) and (\ref{eq:j22}) can be obtained from the solution (\ref{eq:jsolution}) of the linear-response equations.

To find the quadratic-in-$E$ charge currents in N, we again solve the circuit equations (\ref{eq:nmspinrelation})--(\ref{eq:fspinperp}), but for Fourier components at frequency $\Omega = 0$ and $\Omega = \pm 2 \omega$ instead of $\omega$, without the source term proportional to the applied electric field in Eq.~(\ref{eq:nmspinrelation}), but with an additional term in Eq.~(\ref{eq:jconservation}), which accounts for the quadratic-in-$E$ contribution to the longitudinal spin current at the F$|$N interface,
\begin{align}
  \label{eq:jconservation2}
\begin{split}
  \vj_{{\rm s},\Omega}^z(\Omega) =&\,
  \left( j_{{\rm se}\parallel,\Omega}^z + j_{{\rm sm}\parallel,\Omega}^z
  + j_{{\rm s}\parallel,\Omega}^{z(2)} \right) \vm^{\rm eq} \\ &\, \mbox{}
  +
  j_{{\rm sm}\perp,\Omega}^z \ve_{\perp}
  + j_{{\rm sm}\perp,\Omega}^{z*} \ve_{\perp}^*
  .
\end{split}
\end{align}
Solving the circuit equations, and using that $2 |\ve_{\perp} \cdot \ve_y|^2 = 1 - (m^{\rm eq}_y)^2$, gives
\begin{align}
  \delta \bar i_{\Omega}^{x(2)} =&\
  \frac{\sigma_{\rm N}}{d_{\rm N}}
  r_{\Omega} m^{\rm eq}_y \left(1- (m^{\rm eq}_y)^2\right) E^2, \\
  \delta \bar i_{\Omega}^{y(2)} =&\,
  - \frac{\sigma_{\rm N}}{d_{\rm N}}
  r_{\Omega} m^{\rm eq}_x \left(1- (m^{\rm eq}_y)^2\right) E^2.
\end{align}
Here, $r_{\Omega}$ is a response coefficient with the dimension of $\mbox{[length]}/\mbox{[electric field]}$,
\begin{align}
  r_{0} =&\,
  - \frac{\hbar  \theta_{\rm SH}^3 \sigma_{\rm N}^2}{4 e^3 \omega} 
  \nonumber \\ &\, \mbox{} \times
  Z_{\parallel}'(0) Z_{\rm N}^2
  \left[
  \frac{\mbox{Re}\, Z_{{\rm F}\perp}^{\rm m}(\omega)}{|Z_{\perp}(\omega)|^2} + 
  \frac{ \mbox{Re}\, Z_{{\rm F}\perp}^{\rm m}(-\omega)}{|Z_{\perp}(-\omega)|^2} \right], \\
  r_{\pm 2 \omega} =&\,
  - \frac{\hbar  \theta_{\rm SH}^3 \sigma_{\rm N}^2}{8 e^3 \omega} 
  \nonumber \\ &\, \mbox{} \times
  Z_{\parallel}'(\pm 2 \omega) Z_{\rm N}^2 
  \frac{Z_{{\rm F}\perp}^{\rm m}(\pm \omega)
   + Z_{{\rm F}\perp}^{\rm m}(\mp \omega)^*}
  {Z_{\perp}(\pm \omega) Z_{\perp}(\mp\omega)^*} ,
\end{align}
where $Z_{\parallel}'(\Omega)$ is the longitudinal parallel spin impedance,
\begin{equation}
  \frac{1}{Z_{\parallel}'(\Omega)} =
  \frac{1}{Z_{\rm N}}
  + \frac{1}{Z_{{\rm F}\parallel}^{\rm e} + Z_{{\rm FN}\parallel}^{\rm e}} 
  + \frac{1}{Z_{{\rm F}\parallel}^{\rm m}(\Omega) + Z_{{\rm FN}\parallel}^{\rm m}}.
  \label{eq:Zparallel}
\end{equation}
The second and third term in Eq.~(\ref{eq:Zparallel}) account for the relaxation of the quadratic-in-$E$ spin current source $j^{z(2)}_{{\rm sm}\parallel,\Omega}$ into F via spin currents carried by conduction electrons and incoherent, thermal magnons, respectively.

\begin{figure}
\includegraphics[]{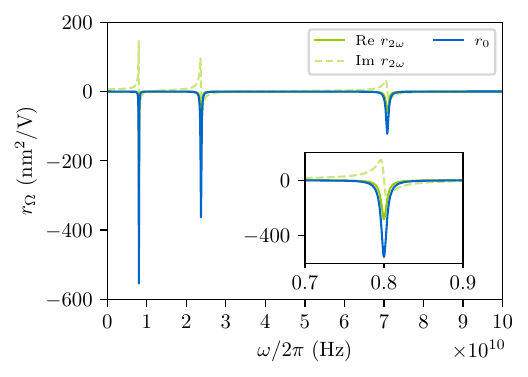}
\caption{\label{fig:numerics1}
Nonlinear response coefficients $r_0$, $\mbox{Re}\, r_{2 \omega}$, and $\mbox{Im}\, r_{2 \omega}$ vs.\ driving frequency $\omega$ for an Au$|$Fe bilayer. Material and device parameters are taken from Ref.\ \onlinecite{Reiss2021-em}, reproduced as Tab.\ \ref{tab:estimates} in the appendix.}
\end{figure}

\begin{figure}
\includegraphics[]{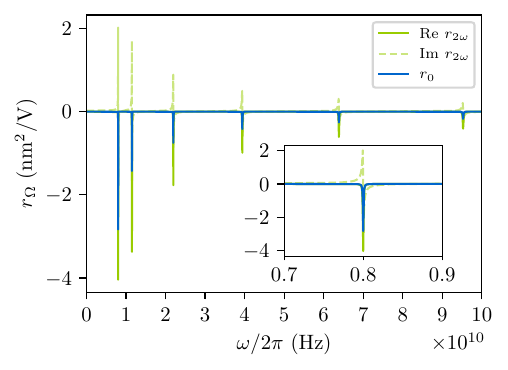}
\caption{\label{fig:numerics2}
Same as Fig.\ \ref{fig:numerics1}, but for a Pt$|$YIG bilayer. Material and device parameters are taken from Ref.\ \onlinecite{Reiss2021-em}, reproduced as Tab.\ \ref{tab:estimates} in the appendix.}
\end{figure}

{\em Numerical estimates for Au$|$Fe and Pt$|$YIG bilayers.---}
In Figs.\ \ref{fig:numerics1} and \ref{fig:numerics2} we show the quadratic response coefficients $r_0$ and $r_{2 \omega}$ as a function of the driving frequency $\omega$ for typical material device parameters for an Au$|$Fe bilayer and for a Pt$|$YIG bilayer. (We regard the ferrimagnet yttrium iron garnet (YIG) as ferromagnetic.) The material and device parameters are the same as in Ref.\ \onlinecite{Reiss2021-em}. They are reproduced in the appendix.

The sharp peaks of the transverse spin impedance $Z_{{\rm F}\perp}^{\rm m}(\omega)$ for driving frequency $\omega$ near the resonance frequencies $\omega_n$, see Eq.~(\ref{eq:omegan}), reflect the spin-torque magnetic resonance. The resonantly driven magnetization modes cause sharp resonances in the unidirectional quadratic-in-$E$ current response. This is the resonant current-in-plane spin-torque diode effect.

The unidirectional response for the Fe$|$Au bilayer is up to two orders of magnitude larger than for YIG$|$Pt, which can mainly be attributed to the longer spin diffusion length $\lambda_{\rm N}$ in Au. Notably, the longitudinal spin transport channels in F have a significant impact on the magnitude of the spin-torque diode effect for metallic ferromagnets and must therefore be considered in a quantitative analysis. These channels are negligible for magnetic insulators because of the absence of spin currents carried by conduction electrons through the interface.

{\em Discussion.---} In this article, we used magneto-electric circuit theory to calculate the spin-torque diode effect in a normal metal--magnet (N$|$F) bilayer. Our theory takes full account of the accumulation of spin at and the flow of spin angular momentum through the N$|$F interface. This is especially important if the magnet is metallic, so that the N$|$F interface can sustain a large spin current carried by conduction electrons.

Our calculation exclusively considered the longitudinal spin currents associated with a resonantly excited coherent magnetization mode. There are additional contributions to the resonant spin-torque diode effect that we did not consider here: a quadratic-in-$E$ contribution to the transverse spin current through the N$|$F interface similar to Eq.~(\ref{eq:j2long}), as well as quadratic-in-$E$ corrections to the spin current carried by thermal magnons and conduction electrons associated with coherent magnetization dynamics. A complete theory including all these contributions to the spin-torque diode effect will be the subject of a forthcoming publication. An additional contribution to the resonant quadratic-in-$E$ response comes from magnetization dynamics driven by the Oersted field \cite{Chiba2014-mh,Chiba2015-jt}. Since the Oersted field scales proportional to the total current in N, this contribution can be made small in the limit of small thickness $d_{\rm N}$ of the normal-metal layer N. We believe our results provide valuable insights for and in the analysis of spin-torque ferromagnetic resonance experiments, in particular for bilayers with a ferromagnetic metal.

{\em Acknowledgments.---}
We thank D. A. Reiss and T. Kampfrath for stimulating discussions. This work was funded by the Deutsche Forschungsgemeinschaft (DFG, German Research Foundation) through the Collaborative Research Center SFB TRR 227 ``Ultrafast spin dynamics'' (Project-ID 328545488, project B03).

\begin{appendix}
\section*{Appendix}

{\em Longitudinal spin impedances.---} We here summarize the expressions for the longitudinal spin impedances from Ref.\ \onlinecite{Reiss2021-em}. We refer to Ref.\ \onlinecite{Reiss2021-em} for a derivation.

The impedance $Z_{{\rm F}\parallel}^{\rm e}$ is given by an expression analogous to Eq.~(\ref{eq:zn}), but with the conductivity $\sigma_{\rm F}$ and spin diffusion length $\lambda_{\rm F}$ of F,
\begin{equation}
  Z_{{\rm F}\parallel}^{\rm e} = \frac{2 e^2}{\hbar}
  \frac{2 \lambda_{{\rm F}}}{\sigma_{{\rm F}}}.
\label{eq:zf}
\end{equation}
(The limit $d_{\rm F} \gg \lambda_{\rm F}$ is assumed here.) The interfacial impedance $Z_{{\rm FN}\parallel}^{\rm e}$ reads
\begin{equation}
  Z_{{\rm FN}\parallel}^{\rm e} =
  \frac{2 \pi (g_{\uparrow\uparrow} + g_{\downarrow\downarrow})}{g_{\uparrow\uparrow} g_{\downarrow\downarrow}}
\end{equation}
where $g_{\uparrow\uparrow}$ and $g_{\downarrow\downarrow}$ are the interfacial conductivities for majority and minority electrons, respectively \cite{Tserkovnyak2005-dt}.

The impedances for spin transport by incoherent, thermal magnons depend on the density of states $\nu_{\rm m}(\varepsilon)$ of magnons with energy $\varepsilon$. Reference \onlinecite{Reiss2021-em} obtains the magnon dispersion from the Landau-Lifshitz equation (\ref{eq:llg}) (without the Gilbert damping term), which gives
\begin{equation}
  \nu_{\rm m}(\varepsilon) = \frac{1}{4 \pi^2} \sqrt{\frac{\varepsilon - \hbar \omega_0}{(\hbar \Dex)^3}}.
\end{equation}
For the interfacial spin impedance, one then has \cite{Bender2015-tr,Reiss2022-tm}
\begin{equation}
  \frac{1}{Z_{{\rm FN}\parallel}^{\rm m}} =
  \frac{1}{\pi s} \mbox{Re}\, g_{\uparrow\downarrow}
  \int d\varepsilon \nu_{\rm m}(\varepsilon) \varepsilon \left( - \frac{df^0}{d\varepsilon} \right),
\end{equation}
where $s$ is the spin per unit volume and $f^0(\varepsilon) = 1/(e^{\varepsilon/k_{\rm B} T} - 1)$ is the Planck function. The spin impedance $Z^{\rm m}_{{\rm F}\parallel}$ reads
\begin{equation}
  Z^{\rm m}_{{\rm F}\parallel}(\omega) =
  \frac{\tau_{\rm rel}}{C_{\rm m}(1 - i \omega \tau_{\rm rel})},
\end{equation}
with
\begin{equation}
  C_{\rm m} = d_{\rm F} \int d\varepsilon \nu_{\rm m}(\varepsilon)
  \left( - \frac{d f^0(\varepsilon)}{d \varepsilon} \right)
\end{equation}
the ``magnon capacity'' and $\tau_{\rm rel}$ the magnon decay time from relativistic non-spin-conserving relaxation processes \cite{Reiss2021-em}
\begin{equation}
  \tau_{\rm rel} = \frac{1}{3 \zeta(3/2) \alpha} \sqrt{\frac{\pi \hbar}{k_{\rm B} T \omega_0}},
\end{equation}
where $\alpha$ is the Gilbert damping constant and $\zeta(3/2) \approx 2.61$.

\begin{table}[t]
\centering
\begin{tabular*}{\columnwidth}{@{\extracolsep{\fill}}ll}
\hline
\hline
 & material and device parameters \\
\hline
Pt & $\sigma_{\rm N} = 9 \cdot 10^6 \, \Omega^{-1} \rm{m}^{-1}$, \\
& $\theta_{\rm SH} = 0.1 $, $\lambda_{\rm N} = 2 \cdot 10^{-9}\, {\rm m}$, \\
& $d_{\rm N} = 4 \cdot 10^{-9}\, {\rm m}$ \\ \hline
Au & $\sigma_{\rm N} = 4 \cdot 10^7 \, \Omega^{-1} \rm{m}^{-1}$, \\
& $\theta_{\rm SH} = 0.08$, $\lambda_{\rm N} = 6 \cdot 10^{-8} \, \rm{m}$, \\
& $d_{\rm N} = 6 \cdot 10^{-8} \, \rm{m}$ \\ \hline
YIG & $\omega_0/2 \pi = 8 \cdot \rm{10^{9} \, Hz}$, $\alpha = 2 \cdot  10^{-4}$, \\
& $\Dex = 8 \cdot 10^{-6} \, \rm{m}^2 \, \rm{s}^{-1}$, $M_{\rm s} = 1 \cdot 10^5 \, \rm{A \, m}^{-1}$, \\
& $d_{\rm F} = 6 \cdot 10^{-8}\, {\rm m}$ \\ \hline
Fe & $\omega_0/2 \pi = 8 \cdot \rm{10^{9} \, Hz}$, $\alpha =  5 \cdot  10^{-3}$, \\
& $\Dex =  4 \cdot 10^{-6} \, \rm{m}^2 \, \rm{s}^{-1}$, $M_{\rm s} = 2 \cdot 10^6 \, \rm{A \, m}^{-1}$, \\
& $\sigma_{\rm F} =  1 \cdot 10^7 \, \Omega^{-1} \rm{m}^{-1}$, \\
& $\lambda_{\rm F} = 9 \cdot 10^{-9} \, \rm{m}$, \\
& $d_{\rm F} = 2 \cdot 10^{-8} \, \rm{m}$ \\ \hline
YIG$|$Pt & $(e^2/h) \textrm{Re} \, g_{\uparrow \downarrow} = 6 \cdot 10^{13} \, \Omega^{-1} \rm{m}^{-2} $, \\
YIG$|$Pt & $(e^2/h) \textrm{Im} \, g_{\uparrow \downarrow} = 0.3 \cdot 10^{13} \, \Omega^{-1} \rm{m}^{-2} $ 
\\
\hline
Fe$|$Au  & $(e^2/h) \textrm{Re} \, g_{\uparrow \downarrow} = 1 \cdot 10^{14} \, \Omega^{-1} \rm{m}^{-2} $ \\
(clean) & $(e^2/h) \textrm{Im} \,  g_{\uparrow \downarrow} = 0.05  \cdot 10^{14} \, \Omega^{-1} \rm{m}^{-2} $, \\
 & $(e^2/h) g_{\uparrow \uparrow} = 4 \cdot 10^{14} \, \Omega^{-1} \rm{m}^{-2}$,\\
& $(e^2/h) g_{\downarrow \downarrow} = 0.8 \cdot 10^{14} \, \Omega^{-1} \rm{m}^{-2}$ \\
\hline
\hline
\end{tabular*}\medskip

\caption{Material and device parameters of the F$|$N bilayers used for the numerical estimate of Figs.\ \ref{fig:numerics1} and \ref{fig:numerics2}, taken from Ref.\ \onlinecite{Reiss2021-em}.}
\label{tab:estimates}
\end{table}

{\em Material and device parameters.---} We take material and device parameters from literature values collected in Ref.\ \onlinecite{Reiss2021-em}. For completeness, the relevant table from Ref.\ \onlinecite{Reiss2021-em} is reproduced here as Tab.\ \ref{tab:estimates}.

\end{appendix}

\typeout{}
\bibliographystyle{apsrev4-2}
\bibliography{main}

\end{document}